\documentclass{MITcsail}

\newcommand{\datasetcurator}{\textit{Dataset Curator}}
\newcommand{\datasetuser}{\textit{Dataset User}}
\newcommand{\datahost}{\textit{Data Host}}
\newcommand{\dataowner}{\textit{Data Owner}}

\usepackage{booktabs}
\usepackage{multirow}
\usepackage[table]{xcolor}
\usepackage{tcolorbox}
\usepackage{array}
\usepackage{subcaption}
\usepackage{fontawesome5}
\usepackage[numbers]{natbib}
\usepackage{amsthm}
\usepackage{lipsum}
\usepackage{xspace}
\usepackage{pgf-pie}
\usepackage{tikz}
\usepackage{textcomp}

\newtcbox{\chipbox}[1][]{nobeforeafter, colback=gray!20, colframe=gray!50,
  boxrule=0.4pt, arc=3pt, boxsep=0pt, left=4pt, right=4pt, top=1pt, bottom=1pt, 
  tcbox raise base, #1}

\definecolor{mint}{RGB}{127,212,198}
\definecolor{yellowpale}{RGB}{255,255,191}
\definecolor{lavender}{RGB}{190,186,218}
\definecolor{coral}{RGB}{251,128,114}
\definecolor{bluepale}{RGB}{128,177,211}
\definecolor{orangepale}{RGB}{253,180,98}
\definecolor{greenpale}{RGB}{179,222,105}

\usepackage[framemethod=tikz]{mdframed}

\definecolor{dimgray}{gray}{0.4}

\newmdenv[
  linecolor=dimgray,     
  linewidth=3pt,         
  topline=false,         
  bottomline=false,      
  rightline=false,       
  leftmargin=0pt,        
  rightmargin=0pt,
  skipabove=10pt,        
  skipbelow=0pt,        
  innertopmargin=4pt,
  innerbottommargin=4pt,
  innerleftmargin=8pt,   
  innerrightmargin=4pt,
  font=\itshape\bfseries
]{takeawaybox}

\title{How Do Data Owners Say No? A Case Study of Data Consent Mechanisms in Web-Scraped Vision-Language AI Training Datasets}

\author{%
  Chung Peng Lee\textsuperscript{\rm 1}\footnote{Correspondence: cl6486@princeton.edu},
  Rachel Hong\textsuperscript{\rm 2},
  Harry H. Jiang\textsuperscript{\rm 3},
  Aster Plotnik\textsuperscript{\rm 4},
  William Agnew\textsuperscript{\rm 3},
  Jamie Morgenstern\textsuperscript{\rm 2,5}\\[1em]
  \normalfont{\small
    \textsuperscript{\rm 1}Princeton University \quad
    \textsuperscript{\rm 2}University of Washington \quad
    \textsuperscript{\rm 3}Carnegie Mellon University \quad
    \textsuperscript{\rm 4}University of Toronto \quad
    \textsuperscript{\rm 5}Amazon AWS AI/ML}%
}

\begin{document}

\maketitle
\thispagestyle{firstpagestyle} 

\begin{abstract}
The internet has become the main source of data to train modern text-to-image or vision-language models, yet it is increasingly unclear whether web-scale data collection practices for training AI systems adequately respect data owners' wishes. Ignoring the owner's indication of consent around data usage not only raises ethical concerns but also has recently been elevated into lawsuits around copyright infringement cases. In this work, we aim to reveal information about data owners' consent to AI scraping and training, and study how it's expressed in DataComp, a popular dataset of 12.8 billion text-image pairs. We examine both the \textit{sample-level} information, including the copyright notice, watermarking, and metadata, and the \textit{web-domain-level} information, such as a site's Terms of Service (ToS) and Robots Exclusion Protocol. We estimate at least 122M of samples exhibit some indication of copyright notice in CommonPool, and find that 60\% of the samples in the top 50 domains come from websites with ToS that prohibit scraping. Furthermore, we estimate 9-13\% with 95\% confidence interval of samples from CommonPool to contain watermarks, where existing watermark detection methods fail to capture them in high fidelity. Our holistic methods and findings show that data owners rely on various channels to convey data consent, of which current AI data collection pipelines do not entirely respect. These findings highlight the limitations of the current dataset curation/release practice and the need for a unified data consent framework taking AI purposes into consideration.
\end{abstract}

\section{Introduction}
\label{sec:introduction}
Web-scraped vision-language datasets (VLD) comprising billions of samples have enabled the success of CLIP~\citep{pmlr-v139-radford21a} as well as text-to-image models like Stable Diffusion v1~\citep{rombach2022high}, DALL-E~\citep{ramesh2021zero}, and MidJourney~\citep{midjourney}. However, the reliance on copyrighted material from the web to train foundation text-to-image or vision language models remains the subject of much recent debate, especially in recent lawsuits against OpenAI, Stability AI, and Meta\footnote{\textit{Andersen v. Stability AI}, No. 3:23-cv-00201 (N.D. Cal.), \textit{Getty v. Stability AI} [2025] EWHC 38 (Ch), \textit{Kadrey v. Meta}, Nos. 3:23-cv-03417, 3:24-cv-06893 (N.D. Cal.), \textit{NYT v. Microsoft}, No. 1:23-cv-11195 (S.D.N.Y.)}. While efforts toward transparent use of copyrighted training data have been explored in text-based pre-training datasets~\citep{longpre2024consent, elazar2024whats}, the data consent landscape of web-scraped VLDs remains relatively underexplored, especially as multimodal image-text models become increasingly common.

The shift from the text modality to the image-text modality results in several changes in data consent mechanisms: 
(1) The signals of data consent in image-text samples are heterogeneous, and (2) image content is often delivered via third-party cloud providers, making the practice of tracking data provenance more challenging. Despite these changes, the impact of violating data consent in the vision-language landscape is no less concerning than that in the text-based counterpart, especially as visual artist communities have spoken out about potential economic loss and reputational harm as a result of generative AI systems \citep{jiang2023ai}.

Furthermore, in recent cases involving Anthropic and Meta\footnote{\textit{Kadrey v. Meta} (see \textit{supra.}), Doc. 598 (Partial Summary Judgment), and 
\textit{Bartz v. Anthropic PBC}, 3:24-cv-05417, (N.D. Cal.), Doc. 231 (Partial Summary Judgment)}, although the training on copyrighted material was deemed “fair use,” the alleged collection of content from pirated sources remains contentious and has precluded the dismissal of the case. This decision raises questions around how dataset curation methods gather data in the first place, and whether such sourcing is allowed. In light of the lack of transparency in web-scraped VLD's data consent~\citep{Hardinges2024We}, we aim to \textit{demystify the data consent mechanisms throughout the life cycle of curating, releasing, and using a web-scraped VLD.}

Specifically, we use DataComp's CommonPool~\citep{gadre2023datacomp} as a case study of the web-scraped VLDs. They sourced image-text pairs from CommonCrawl~\citep{commoncrawl}, an archive of web pages crawled from the internet, and performed deduplication and minimal filtering to produce a set of 12.8B \textit{url-text} pairs, where the \textit{url} points to the image content. As of July 2025, CommonPool has over 2M downloads~\citep{datacomphuggingface}. Pulling from the same web archive, CommonPool has substantial overlap with its precursor, LAION-5B~\citep{schuhmann2022laion}, which enabled the early version of Stable Diffusion v1, MidJourney, and Google's Imagen~\citep{rombach2022high, midjourney, saharia2022photorealistic}. Even though the data used to train OpenAI's CLIP or DALL-E were not disclosed, the corresponding papers claim to have sourced the training datasets from the internet~\citep{pmlr-v139-radford21a, ramesh2021zero}, similar to CommonPool. Therefore, we believe CommonPool as a case study not only informs the open-source vision-language model development community but also provides a lens into commercially protected datasets.

We recognize and take advantage of various signals provided by the image, text, metadata, and their associated data host. We use both sample-level characteristics, such as copyright notice, the exchangeable image file format (EXIF)~\footnote{https://en.wikipedia.org/wiki/Exif} metadata, and watermark detection, and web-domain-level characteristics, such as Terms of Service (ToS) and Robots Exclusion Protocols (REP), also known as robots.txt. We make the following contributions:
\begin{enumerate}
    \item Investigate data consent mechanisms in a web-scraped VLD provided by the information in the released artifact
    \item Estimate approximately 122M of samples in CommonPool have included copyright information, and over 60\% of samples from the top 50 domains, in the \texttt{small-en} scale of CommonPool, are sourced from sites restricting scraping in their ToS.
    \item Demonstrate that data owners often rely on inconsistent channels to convey data consent, of which AI data collection pipelines do not fully respect, surfacing issues of a lack of a uniform consent mechanism.
    \item Use our findings to outline various limitations and recommendations for future web-scraped VLD curation.
\end{enumerate}
\section{Background}
\label{sec:background}
\subsection{Terminology}
In this section, we outline the scope of each term and the role they play in the explicit permission granted to use the data. We limit our focus to examining data consent and copyright implications within the United States.
\subsubsection{Copyright} As defined by the U.S. Copyright Office~\citep{copyright-office}, copyright protects the expression of original work. As long as the work is \textit{fixed}, \textit{expressed in tangible forms}, and not an idea, concept, fact, or other exception, it automatically becomes copyright-protected. Notably, the role of the \textit{copyright notice}, like ``© John Doe 2025'', is to publicly claim that the work is protected by copyright. As such, it becomes more difficult for defendants in infringement cases to argue they were not aware of the work being copyrighted~\citep{copyright-notice}.
\subsubsection{License} A license, or agreement, grants specified rights to someone to use the work for purposes protected by copyright, such as reproduction, display, or making derivatives. A license could be useful for the creator to limit the use of the work in certain scenarios without placing it in the \textit{public domain}, which is outside the scope of copyright protection.
\subsubsection{Data Consent} We refer to data consent as the ``permission'' granted for the user to use the data for model training purposes. This is not limited to any form of written consent, such as ToS, copyright notice, claims, or license. In other words, data consent is obtained when the user follows the acceptable pipeline to retrieve data proposed by the data host or data owner. As an example, even if the data is not copyright-registered through the U.S. Copyright Office, a written ToS to restrict the use of such data for model training purposes would be considered a ``restriction to use'' in the scope of data consent we consider.

\subsection{Involved Parties}
The pipeline to curate, release, and download a web-scraped dataset involves multiple entities. To study the data consent landscape, we first define how the stakeholders are involved in the life cycle of such datasets.

\begin{itemize}
    \item \datasetcurator{} -- The curator of the dataset releases a set of \textit{url-text} pairs for downstream use. In the case of DataComp~\citep{gadre2023datacomp}, it would be their authors.
    \item \datasetuser{} -- The user of the dataset downloads the pairs of URLs and texts released by the \datasetcurator{}.
    \item \dataowner{} -- The owner of the image data itself. Since tracing data ownership on the internet is extremely difficult, we relax the ownership to be the action of embedding the image on their web page. This relaxation builds on the assumption that the actor of embedding the image respects the copyright of the image and shares it per the level of consent they obtain.
    \item \datahost{} -- The data host is the entity that owns the image URL referred to by the sample. Since the delivery of image content is often optimized through content delivery network (CDN) and cloud providers, this entity may exhibit little information about the \dataowner{.}
\end{itemize}

\begin{figure*}[!t]
    \centering
    \includegraphics[width=\textwidth]{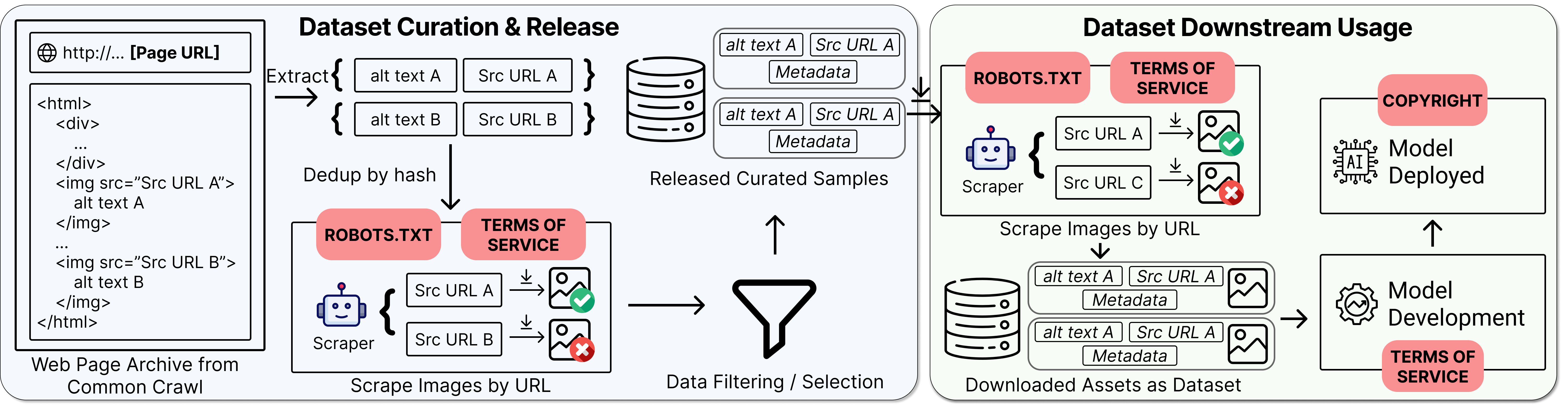}
    \caption{
    The life cycle of curating, releasing, and using the web-scraped VLD. Even though the \datasetcurator{} initially downloads the image assets in their curation process, the released samples only contain the caption, \textit{src url} pointing to the image asset, and image metadata. To access the dataset, the \datasetuser{} must download the images following the released URLs. The red tags on each step indicate the data consent mechanism we consider involved.}
    \label{fig:consent-table-figure}
\end{figure*}

\subsection{Life Cycle of Web-Scraped VLD}
\subsubsection{Curation \& Release}
The top-level raw source of data originates from CommonCrawl~\citep{commoncrawl}. The collection of \textit{url-text} pairs comes from extracting the \texttt{<img src=URL>alt text</img>} from the internet. This extraction \textit{does not} consider the \textit{page url} where the image appears. Figure \ref{fig:consent-table-figure} illustrates the distinction between \textit{page url} and \textit{src url}. With the extracted \textit{url-text} pairs, the \datasetcurator{} uses tools like img2dataset~\citep{beaumont-2021-img2dataset} to automatically download all the images from these URLs, referred to as \textit{scraping}. Since the URLs are extracted from archives of the internet, \textit{not all download attempts are successful or align with the original image}. For instance, the owner of the URL could replace the image with another image or take down the image completely. With the downloaded assets, the \datasetcurator{} experiment with data filtering, cleaning, augmentation, and model training/evaluation to curate the best set for release. Finally, the release of the curated dataset comprises \textit{url-text} pairs along with metadata they obtain from either their experiments or downloading, \textit{without the actual image assets.}
\subsubsection{Downstream Usage}
The \datasetuser{} first obtains the index of \textit{url-text} pairs released by the \datasetcurator{}. Since the released dataset artifact comes without the image assets, the \datasetuser{} has to utilize similar tools to \textit{scrape} through the provided URLs. In the case of DataComp~\citep{gadre2023datacomp}, the scraping functionality is provided as part of the release. This mechanism inherits the same drawback of potentially inconsistent or failed downloads. Not only does it potentially diverge from the \datasetuser{'s} expectation of the released dataset, but it might also expose the \datasetuser{} to the risk of data poisoning~\citep{carlini2024poisoning}. Furthermore, since the \datasetuser{} is scraping the web with the index of the URLs, the \datasetuser{} is responsible for abiding by any ToS or other data consent mechanism specified by the website hosting the content. With the image assets downloaded, the \datasetuser{} then experiments with the downloaded samples in their storage.
\section{Methodology}
\label{sec:methodology}
We first outline the concrete experiment setup for our audit, including data filtering, sizes, and scales that we audit. Then, we present the methods in two categories, one at the sample level and the other at the web domain level. These two angles allow us to audit how image owners and website owners disclose consent for scraping and AI training.
\subsection{Setup}
CommonPool was released at four scales: \texttt{xlarge} (12.8B), \texttt{large} (1.28B), \texttt{medium} (128M), and \texttt{small} (12.8M), where the largest contains 12.8B samples and the lower scale is a subset of the larger ones. Due to limited storage space and compute resources, we study both \texttt{small} and \texttt{medium} such that we can verify whether results found in \texttt{small} are also observed in \texttt{medium}.

Moreover, since legal mechanisms of data consent are dependent on specific jurisdictions, we restrict our target data to be English-based. Particularly, we follow the same measure in \citet{gadre2023datacomp} to use \texttt{fasttext}~\citep{joulin2016bag} to filter the original dataset by English-only captions. Table \ref{tab:dataset-summary} summarizes the audited dataset.

\begin{table}
\centering
\begin{tabular}{lccc}
\toprule
\textbf{Scale}    & \textbf{Released} & \textbf{Accessible} & \textbf{``Top 50''} \\ \midrule
\texttt{small}          & 12.8M             & 9.8M         & -- \\
\texttt{small-en}       & 6.3M              & 4.8M         & 2.1M \\
\texttt{medium}         & 128.0M            & 98.3M        & -- \\
\texttt{medium-en}      & 63.0M             & 47.7M        & 21.5M \\
\bottomrule
\end{tabular}
\caption{Sample counts of CommonPool's configurations considered in our work. \texttt{scale-en} refers to the English-filtered version of the original scale. Accessible counts refer to images downloadable through the released link. ``Top 50'' refers to the subset in the top 50 \textit{base domains}.}
\label{tab:dataset-summary}
\end{table}

\subsection{Sample-level Characteristics}
At the sample level, we use text, visual, and metadata information to source characteristics of data consent. Particularly, we search for samples with the presence of \textit{copyright notice}, \textit{copyright field in metadata}, and \textit{image watermark}. With the presence of this information, it is difficult for a defendant on copyright infringement to argue ignorance of the fact that the material was copyright-protected~\citep{copyright-notice}.

\subsubsection{Copyright Notice}
We crafted a set of regular expressions to capture common copyright notices such as ``©'' and ``copr.'' These rules are applied to both caption and OCR-extracted text, where we use open-source PaddleOCR~\citep{paddleocr2020} for extraction. The full list of search patterns is included in Appendix.

\subsubsection{Copyright Field in Metadata} \textit{Exchangeable image file format} (EXIF) is a standard of image metadata to specify information about the image as well as the digital device that produced the image. For instance, some tags include original height, width, and focal length. We search for samples of which the metadata contains a non-empty copyright tag field keyed by ``Copyright'' or ``0x8298,'' following the EXIF standard version 2.3.~\citep{exifstandard}.

\subsubsection{Image Watermark}
A watermark detection classifier aims to output whether or not a given image contains a watermark. We (1) use off-the-shelf watermark-finetuned YoloV8~\citep{ultralytics, watermark-yolo-v8}, (2) build a watermark-finetuned MobileViTv2~\citep{mehta2022separable}, (3) use two open-source VLMs, Rolm OCR~\citep{RolmOCR} and Gemma-3-12b-it~\citep{gemma_2025} as our detection methods. To validate the faithfulness of these methods, we evaluate them on (1) \textit{watermark-eval}: \citet{pollano_watermarked_2019}'s validation set, with a balance of $\sim$3200 images for both watermarked and non-watermarked images, and (2) \textit{datacomp-watermark-eval}: a random 955-image subset of CommonPool we annotate, to validate the robustness of our detection methods on web-scraped images. Last but not least, we question the faithfulness of LAION-5B's release of \textit{watermark score} by annotating a subset of LAION-5B and analyzing the utility of those scores\footnote{LAION-5B releases watermark scores per sample to estimate the probability of the presence of watermark in the image.}. The full training and evaluation details can be found in Appendix.

\subsection{Web-Domain-Level Characteristics}
At the web-domain level, the administrator who hosts the content typically specifies rules on permitted usage of their content. Particularly, we examine the top 50 web domains' ToS and their REP, which specifies the restriction of scraping/crawling bots. The top 50 domains are defined by the counts of samples sourced from these domains. In both \texttt{small-en} and \texttt{medium-en} scales, the top 50 domains cover $\sim$45\% of all samples, namely 2.1M and 21.5M samples respectively.

The web domains are extracted from \textit{src url} as provided by CommonPool, which points to the image asset, rather than the original website where the content is embedded, which we call \textit{page url}. Furthermore, since most content is delivered through domains designed for static content or a content delivery network (CDN), we extract the \textit{base domain} by trimming off the prefix to aggregate the sharded domain URLs. For instance, Pinterest uses bucketed web domains like \texttt{i.pinimg.com} and \texttt{i-h1.pinimg.com} to deliver content. Through extracting only the \textit{base domain}, which would be \texttt{pinimg.com} in the example, we have a more accurate estimate of sample counts for each web domain.

\subsubsection{Terms of Service (ToS)} Following \citet{longpre2024consent}, we annotate each web domain with the following attributes: (1) Category: the core function of the \datahost{}, (2) License Type: the permission granted to the end user, and (3) Scraping Policy: the restriction on web-scraping. In this work, we focus on the act of \textit{scraping}, the action of automatically downloading/copying a vast majority of data through an index of links, because both the \datasetuser{} and \datasetcurator{} directly engage in this act.\footnote{In contrast, \textit{crawling} refers to the act of developing a spider to recursively follow links from web pages to store content.}

Similar to \citet{fiesler2016reality}'s qualitative analysis process, we have two coders to annotate each web domain's attributes, but we start with the codebook for (2) and (3) from \citet{longpre2024consent}. For the Category, the primary coder first builds the codebook when iteratively going through the web domains. After creating the initial codebook and first pass, the second coder annotates the web domains. The two coders resolve any conflict through adjusting either the annotations or the codebook. The types in each attribute and the full codebook are included in the Appendix.

\subsubsection{Robots Exclusion Protocol (REP)}
REP, implemented via robots.txt, allows website administrators to specify which automated clients (user agents) can access their sites. Administrators can allow or disallow access for specific agents, such as ``CCBot'' (CommonCrawl), ``GPTBot'' (OpenAI), or any agent using the wildcard ``*''. They can also restrict access to certain website paths. In Germany, robots.txt is legally enforceable, with exceptions for scientific research~\citep{Hamburg_2024, EU2019dsm}.

For each of the top 50 \textit{base domains}, we map the \textit{base domain} to a list of \textit{full domains}, which are the web domains with the original prefix. For instance, the \textit{base domain}, \texttt{pinimg.com}, maps to a list of \textit{full domains}, \texttt{[i.pinimg.com, i-h1.pinimg.com, ...]}. We retrieve robots.txt by appending ``robots.txt'' at the end of the \textit{full domains}. In the \texttt{small-en} scale, there are 96,436 unique URLs requested, and 81,273 of them successfully return with a non-empty robots.txt\footnote{In the \texttt{medium-en} scale, 434,498 URLs were requested, and 392,286 successfully returned with a non-empty robots.txt.}.

We parse each robots.txt following \citet{longpre2024consent} to three categories: \textit{All Disallowed}, \textit{Some Disallowed}, and \textit{None Disallowed} for agents listed in the robots.txt file. \textit{All Disallowed} is when a particular agent is mentioned and disallowed from all parts of the site. \textit{None Disallowed} is when ``the particular agent is mentioned and allowed for all parts of the site,'' or ``has no disallowed parts.'' \textit{Some Disallowed} is when ``a particular agent is mentioned and disallowed from some parts of the website.'' \textit{Some Disallowed} is when a particular agent is mentioned and disallowed from some parts of the website. An agent must be listed in robots.txt to determine the category.
\section{Results}
\label{sec:results}
In this section, we present our findings according to the sample-level and web-domain-level methods of determining data consent.

\begin{table}[ht]
\centering
\begin{tabular}{lll}
\toprule
\textbf{Measure} & \texttt{small-en} & \texttt{medium-en} \\
\midrule
Caption          & 10,585 (0.22\%)   & 98,555 (0.21\%)     \\
OCR              & 4,307 (0.09\%)    & 38,697 (0.08\%)     \\
EXIF Metadata    & 108,951 (2.27\%)  & 1.09M (2.28\%)  \\
Caption $\cup$ OCR $\cup$ EXIF & 123,096 (2.56\%) & 1.22M (2.55\%) \\
\bottomrule
\end{tabular}
\caption{Number of samples found through each measurement method, where Caption and OCR refer to searching the copyright notice through samples' captions and OCR-extracted texts.}
\label{tab:sample-level-characteristics}
\end{table}

\subsection{Sample-Level Statistics}
\begin{takeawaybox}
    $\mathbf{\mathord{\sim}122\text{M}}$ English samples in CommonPool contain characteristics of copyright notice or claims.
\end{takeawaybox}
We find 1.22M samples exhibiting characteristics of copyright notice or claims in the \texttt{medium-en} scale. We further validate the faithfulness as the portions of the found samples through each method scale similarly from \texttt{small-en} to \texttt{medium-en}, as shown in Table \ref{tab:sample-level-characteristics}. This extends our results to implications on the full dataset of 12.8B samples, where approximately 122M of English samples may contain copyright notices or claims. We observe very little overlap between the keyword search methods across image, text, and EXIF  metadata. This signifies that copyright claims are heterogeneously disclosed for images on the internet, which emphasizes the need to examine each modality to adequately determine copyright information from web-scraped samples.

\begin{table}[!ht]
\centering
\resizebox{\textwidth}{!}{
\begin{tabular}{lcccccccc}
\toprule
\multirow{2}{*}{\textbf{Model}} & \multicolumn{4}{c}{\textit{wm-eval}} & \multicolumn{4}{c}{\textit{datacomp-wm-eval}} \\
 & Accuracy & Precision & Recall & F1 & Accuracy & Precision & Recall & F1 \\
\midrule
Finetuned YoloV8         & 96.69 & 97.44 & 95.90 & 96.66 & 86.91 & 42.63 & 51.88 & 46.80 \\
Finetuned MobileViTv2      & 89.25 & 90.43 & 86.63 & 88.49 & 30.37 & 11.02 & 74.53 & 19.20 \\
Rolm-OCR                 & 74.62 & 99.15 & 49.74 & 66.25 & 89.10 & 50.80 & 59.43 & 54.78 \\
Gemma-3-12b-it           & 90.66 & 99.22 & 81.87 & 89.71 & 85.34 & 41.05 & 73.58 & 52.70 \\
\bottomrule
\end{tabular}
}
\caption{Evaluation of watermark detection methods on both standard watermark detection dataset, \textit{wm-eval} with 3289 clean and 3299 watermark images, and an annotated set of web-scraped images from CommonPool, \textit{datacomp-wm-eval} with 849 clean and 106 watermark images.}
\label{tab:watermark-detection-result}
\end{table}

\begin{takeawaybox}
    Watermarks are present in web-scraped images, but detecting them remains a major challenge — even for conventionally advanced methods.
\end{takeawaybox} 
In our evaluation suites, we use (1) \textit{watermark-eval}, comprising a balance of 3289 clean and 3299 watermarked images, and (2) \textit{datacomp-watermark-eval}, a random sample of 955 images from CommonPool we annotate. We find that 106 of those images, or 11.09\%, are watermarked, resulting in a 9\% to 13\% of the distribution with 95\% confidence interval. From Table \ref{tab:watermark-detection-result}, we observe that across all models, the F1-score significantly drops on \textit{datacomp-wm-eval}. This indicates a distribution shift between the traditional watermark detection dataset and the web-scraped images ``in the wild.'' Upon investigation, we determine that traditional methods tend to have lower precision on \textit{datacomp-watermark-eval} because of the text appearing in the image, where the models tend to output \textit{True} for images with texts in them.

\begin{takeawaybox}
    Is LAION-5B's released watermark score reliable for understanding and respecting data consent?
\end{takeawaybox}
In light of our watermark detection experiments, we question the fidelity of the watermark score released in LAION-5B~\citep{schuhmann2022laion}. We annotate 1308 random samples from LAION-5B and find that 176 have a watermark, or 13.45\%. Furthermore, using the standard threshold of 0.5 on the watermark scores released, the precision and recall are only at 34.09\% and 51.13\%. The area under the receiver operating characteristic (ROC) curve is 0.74. These statistics further demonstrate the difficulty of watermark detection for web-scraped images ``in the wild'' observed in our experiments. Moreover, the low performance of LAION-5B's watermark score reveals the low utility of this watermark probability score if a dataset user wishes to avoid training AI systems on watermarked images.

\subsection{Web-Domain-Level Statistics}
Since the top 50 base domains in \texttt{small-en} and \texttt{medium-en} only differ by 1 base domain, we present the results for \texttt{small-en} for conciseness. The distribution of the top 50 base domains can be found in the Appendix. For robots.txt, we primarily present our results with the top six user agents in terms of the number of ``observations,'' or samples that come from sites with robots.txt files that mention the top six agents. The total number of observed agents, weighted by sample counts, is 1.1M. Full results are included in Appendix.

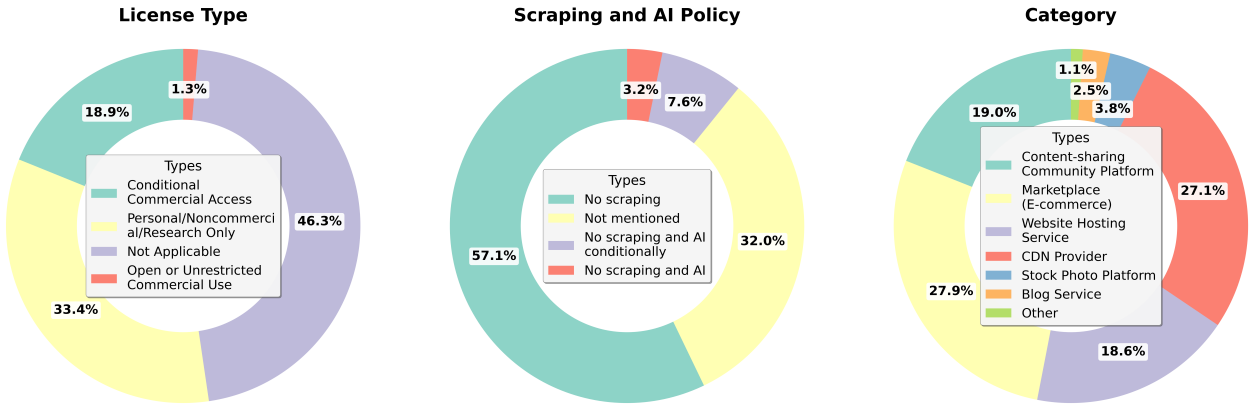
\begin{figure}[ht]
\newcommand{\ToSPieChartsSep}{7}

\centering
\resizebox{\textwidth}{!}{%
\begin{tikzpicture}[noborder/.style={draw opacity=0}]

\begin{scope}[shift={(-\ToSPieChartsSep,0)}]
    \node[font=\bfseries] at (0,3.3) {License Type};
    \pie[
        pos={0,0},
        radius=2.8,
        text=inside,
        color={mint, yellowpale, lavender, coral},
        style=noborder,
        before number={},
        after number={\%}
    ]{
        18.9/,
        33.4/,
        46.3/,
        1.3/
    }
    \node [font=\scriptsize, anchor=north] at (0,-3.2) {
        \begin{tabular}{ll}
            \tikz\fill[mint] (0,0) rectangle (0.25,0.12); & Conditional Commercial Access \\
            \tikz\fill[yellowpale] (0,0) rectangle (0.25,0.12); & Personal/Noncommercial/Research Only \\
            \tikz\fill[lavender] (0,0) rectangle (0.25,0.12); & Not Applicable \\
            \tikz\fill[coral] (0,0) rectangle (0.25,0.12); & Open or Unrestricted Commercial Use
        \end{tabular}
    };
\end{scope}

\begin{scope}[shift={(0,0)}]
    \node[font=\bfseries] at (0,3.3) {Scraping and AI Policy};
    \pie[
        pos={0,0},
        radius=2.8,
        text=inside,
        color={mint, yellowpale, lavender, coral},
        style=noborder,
        before number={},
        after number={\%}
    ]{
        57.2/,
        32.0/,
        7.6/,
        3.2/
    }
    \node [font=\scriptsize, anchor=north] at (0,-3.2) {
        \begin{tabular}{ll}
            \tikz\fill[mint] (0,0) rectangle (0.25,0.12); & No scraping \\
            \tikz\fill[yellowpale] (0,0) rectangle (0.25,0.12); & Not mentioned \\
            \tikz\fill[lavender] (0,0) rectangle (0.25,0.12); & No scraping and AI conditionally \\
            \tikz\fill[coral] (0,0) rectangle (0.25,0.12); & No scraping and AI
        \end{tabular}
    };
\end{scope}

\begin{scope}[shift={(\ToSPieChartsSep,0)}]
    \node[font=\bfseries] at (0,3.3) {Category};
    \pie[
        pos={0,0},
        radius=2.8,
        text=inside,
        color={mint, yellowpale, lavender, coral, bluepale, orangepale, greenpale},
        style=noborder,
        before number={},
        after number={\%}
    ]{
        19.0/,
        27.9/,
        18.6/,
        27.1/,
        3.8/,
        2.5/,
        1.1/
    }
    \node [font=\scriptsize, anchor=north] at (0,-3.2) {
        \begin{tabular}{ll}
            \tikz\fill[mint] (0,0) rectangle (0.25,0.12); & Content-sharing Platform \\
            \tikz\fill[yellowpale] (0,0) rectangle (0.25,0.12); & Marketplace \\
            \tikz\fill[lavender] (0,0) rectangle (0.25,0.12); & Website Hosting \\
            \tikz\fill[coral] (0,0) rectangle (0.25,0.12); & CDN Provider \\
            \tikz\fill[bluepale] (0,0) rectangle (0.25,0.12); & Stock Photo Platform \\
            \tikz\fill[orangepale] (0,0) rectangle (0.25,0.12); & Blog \\
            \tikz\fill[greenpale] (0,0) rectangle (0.25,0.12); & Other
        \end{tabular}
    };
\end{scope}

\end{tikzpicture}%
}
    \caption{Terms of Service annotations. The full population in each chart is all samples in the top 50 base domains of \texttt{small-en}. The portion is determined by the exact number of samples in each type. For License Type, ``Not Applicable'' indicates that the ToS from the base domain does not specify or provide any license type information. For Category, ``Other'' indicates that the base domain is for a very domain-specific service. For instance, \textit{4sqi.net} is delivered by Foursquare, a location-intelligence service provider.}
    \label{fig:ToS-analysis}
\end{figure}

\begin{takeawaybox}
    60\% of samples in the top 50 base domains prohibit scraping, and 33\% of them are restricted to Personal / Research / Non-commercial Only Use.
\end{takeawaybox}

Through our analysis of the ToS in Figure~\ref{fig:ToS-analysis}, 57.1\% of the top 50 base domains prohibit general scraping without mentioning AI, and 3.2\% prohibit scraping and AI unconditionally. This not only emphasizes the responsibility of the \datasetcurator{} but also that of the \datasetuser{}, who scrapes these sites as well while downloading CommonPool. Furthermore, 33.4\% of samples in the top 50 base domains come from websites with ToS limiting usage of content for Personal/Research/Non-commercial purposes.



\begin{takeawaybox}
    Releasing only url-text pairs restricts the ability to examine data consent through ToS.
\end{takeawaybox}
Web-scraped VLDs, such as CommonPool, LAION-400M, and LAION-5B, all use the practice of releasing only the \textit{src url} and caption as described in Section~\ref{sec:background}. We find that 27.1\% and 18.6\% of the samples in the top 50 base domains are under CDN Provider and Website Hosting Service categories, respectively. Yet, the ToS of \texttt{amazonaws.com} cannot fully reflect the actual ToS used by the website offering the content stored at those \textit{src urls}. The core reason is that image content delivered via \textit{src url} is often through a CDN or static content host, and only those \textit{src urls} are released instead of the original \textit{page url}. Without the context of \textit{page url}, the website URL where the \textit{url-text} pair is extracted, a thorough examination of data consent is infeasible. This characteristic also primarily accounts for the reason why 46.9\% of samples' License Type in Figure~\ref{fig:ToS-analysis} are categorized as ``Not Applicable,'' meaning that the provided \textit{src urls}' base domain's ToS may not have the right to specify the License Type.

\begin{takeawaybox}
    Robots.txt is predominantly adopted to convey restrictions for AI-purpose scrapers/crawlers.
\end{takeawaybox}

In the top 6 agents by number of samples covered by observations, we see that traditional web-indexing (googlebot-image) or wildcard ($\ast$) agents don't have very high \textit{All Disallowed} rate compared to agents related to AI-purposes such as GPTBot, Bytespider, and claudebot. This phenomenon implies that the website administrator disallowing these AI-purpose agents wishes to prevent the use of their content for model development. However, a dataset user downloading CommonPool to train a model does not specify the user agent by default and therefore can bypass REP to scrape many of these same samples from sites that ban GPTBot, Bytespider, and claudebot. Only 3.9\% of samples come from sites that disallow any agent, so many sites that specifically block AI-purpose bots may miss dataset users scraping open-source VLDs to train models. 

\begin{takeawaybox}
    The temporal shift in data consent may not be reliably reflected in scraping.
\end{takeawaybox}
Moreover, even though CommonPool is sourced from CommonCrawl, which respects robots.txt when sourcing the web pages, we still observe CCBot in 353K robots.txt. We hypothesize that the user adopts robots.txt to revoke their consent after CommonCrawl archives their pages. Despite this adoption, the collection of CommonPool as an index of url-text pairs continues to direct scraping traffic to those websites that chose to revoke consent when the \datasetuser{} downloads CommonPool using a non-CCBot user agent name.

\begin{table}[ht]
\centering
\resizebox{\textwidth}{!}{
\begin{tabular}{lc cc cc cc}
\toprule
\multirow{2}{*}{\textbf{Agent}} & \multirow{2}{*}{Observed} 
& \multicolumn{2}{c}{\textit{All Disallowed}} 
& \multicolumn{2}{c}{\textit{Some Disallowed}} 
& \multicolumn{2}{c}{\textit{None Disallowed}} \\
 &  & Count & \% of observed & Count & \% of observed & Count & \% of observed \\
\midrule

``All Agents'' & 1,126,876 & 6,442 & 0.6\% & 1,014,576 & 90.0\% & 105,858 & 9.4\% \\ \midrule
\rowcolor{gray!20} GPTBot \faRobot            & 578,498  & 538,431 & 93.1\% & 40,028 & 6.9\% & 39 & 0.0\% \\
$\ast$            & 475,139  & 18,595  & 3.9\%  & 391,799 & 82.5\% & 64,745 & 13.6\% \\
\rowcolor{gray!20} CCBot \faRobot             & 353,324  & 313,920 & 88.8\% & 39,365 & 11.1\% & 39 & 0.0\% \\
\rowcolor{gray!20} Bytespider \faRobot        & 301,344  & 262,029 & 87.0\% & 39,274 & 13.0\% & 41 & 0.0\% \\
googlebot-image   & 224,268  & 0       & 0.0\%  & 224,166 & 100.0\% & 102 & 0.0\% \\
\rowcolor{gray!20} claudebot \faRobot         & 224,200  & 224,199 & 100.0\% & 1 & 0.0\% & 0 & 0.0\% \\
\bottomrule
\end{tabular}
}
\caption{Top results from robots.txt analysis for \texttt{small-en} scale's top 50 \textit{base domains}, accounting for 96,436 attempted \textit{full domains}, 81,273 successful robots.txt, and 1,126,876 samples observed. For each agent, the number of observed cases is broken down by the number and percentage (relative to observed) of cases where all, some, or none were disallowed. The dark gray background highlights rows that have over 80\% \textit{All Disallowed} rate, and the \faRobot{} icon indicates that the agent is AI-purposed. ``All Agents'' row refers to an aggregation of all agents found in all the examined robots.txt. The aggregation rule is as follows: If for all agents, a robots.txt has \textit{All Disallowed}, then the decision is \textit{All Disallowed}. If for any agent in all agents, a robots.txt has \textit{All Disallowed} or \textit{Some Disallowed}, then a robots.txt has \textit{Some Disallowed}. Otherwise, it has \textit{None Disallowed}.}
\label{tab:agent-disallowed-summary}
\end{table}

\section{Discussion}
\label{sec:discussion}

 
\subsection{Limitation of Current Release Practice}
\subsubsection{Problem} Our results reveal several drawbacks in the current release practice of web-scraped VLDs. Firstly, the lack of \textit{page url} greatly restricts the ability to probe whether an image is prohibited from use by the associated ToS. This issue originates from a combination of how image content is usually delivered through CDN, how each sample is collected by only an HTML tag, and how the website itself (\textit{page url}) is not always related to the extracted HTML tag. Secondly, releasing an index of the web through \textit{url-text} pairs allows the \datasetcurator{} to avoid hosting any image asset, and thus any copyright infringement claim or responsibility of providing a convenient channel for the \datasetuser{} to access the copyrighted/restricted-to-use data. This shift of accountability may not be made aware to the \datasetuser{}, creating an illusion that the curation of an open-sourced web-scraped VLD has already dealt with data consent, so usage of that dataset is in the clear.
\subsubsection{Recommendation} For better data provenance and transparency, we recommend that future releases include the website page where the samples are collected. Moreover, the \datasetcurator{} should either \textit{clearly inform or warn} the \datasetuser{} about the potential responsibility of scraping when using their dataset, or take the responsibility to construct the dataset with standalone image assets respecting the \dataowner{}'s consent, through the various mechanisms we used in our audit.
\subsection{Call for a Unified Data Consent Framework}
\subsubsection{Problem} In our case study of DataComp CommonPool, we find that each audit approach surfaced a distinct set of samples restricting data usage with very few overlaps. This observation indicates that the data consent is conveyed through multiple channels, such as image metadata, copyright notice, or image watermark. Even though this highlights the importance of auditing through our comprehensive techniques, it presents a problem of lacking a universally recognized framework to convey data consent, particularly in the life cycle of AI data collection. For instance, robots.txt was constructed for web scraping, but web scraping is only a part of the life cycle. As another example, the copyright notice goes beyond the consent for model development, but also for display, re-distribution, and so on. In addition to the divergent channels to convey data consent, \citet{longpre2024consent} reveals a contradiction between these channels where ToS have different restrictions from REP.
\subsubsection{Recommendation} All the involved parties highlighted in this work need a common protocol such that data owners can communicate data consent, specifically for the use of model development. The Robots Exclusion Protocol is not sufficient because we showed that website maintainers often are not the owners of the data. We believe that a unified channel not only helps the \dataowner{} to protect their works from misuse, but also guides the \datasetcurator{} and \datasetuser{} to respect their data consent. Such a framework should not only be adopted but also treated as the source of truth to represent data consent. In addition, we encourage the adoption of an opt-in understanding of consent, as supported by many data owner stakeholders~\citep{3cs-chi, 3cs-orig}. Existing solutions, e.g. an opt-out model~\citet{spawning}, do not address the obscurity of scraping and training to many data owners, and implicitly obfuscate consent. Recently proposed Human Commons~\citep{humanscommons_2025} can be viewed as a specialized consent mechanism acknowledging the complexity and uniqueness of the problem, but the community adoption is still in progress. In short, although a variety of frameworks have been proposed with their merits, there is still a lack of consensus on which to adopt and which to respect.
\section{Related Work}
\label{sec:related_work}
Prior work on auditing web-scale pre-training datasets ranges from data governance, privacy to social biases. In the text modality, \citet{dodge2021documenting} highlighted the importance of documenting datasets with the excluded data's characteristic, web domain distribution, and other aspects of Colossal Clean Crawled Corpus (C4)~\citep{JMLR:v21:20-074}. \citet{elazar2024whats} extended the goal to understand these datasets to several pre-training datasets, such as C4, LAION-2B-en, and The Pile~\citep{JMLR:v21:20-074, schuhmann2022laion, pile}, by documenting their domain statistics, contamination with evaluation sets, and PII inclusion. More specific to data consent, \citet{longpre2024consent} investigated the consent mechanism of text-based pre-training datasets including C4, dolma, and RefinedWeb~\citep{JMLR:v21:20-074, soldaini2024dolma, penedo2023the}. They focus on the temporal changes in data consent in both ToS and robots.txt and highlight the increasing restrictions on the web to train AI models with web-scraped data.

In the vision-language datasets landscape, \citet{hong2024s} studied the impact of data filtering on the exclusion/inclusion statistics concerning minority groups across gender, religion, and race. \citet{hong2025common} presented a legally-grounded study on private information existing in CommonPool and its implications from a legal perspective. Our work studies the data consent mechanism in the landscape of web-scraped VLDs.
\section{Conclusion}
\label{sec:conclusion}
In this work, we fill the gap to explore the data consent mechanism of web-scraped VLDs. Particularly, we employ various approaches taking advantage of the rich information provided by the data. We not only find that a significant number of samples, projected 122M in CommonPool, can have copyright claims and/or information, but also a huge portion of top-domain samples, around 60\%, are under a scrape-restricted web domain. Furthermore, we shed light on several concerns and limitations of the current curation and release practice of web-scraped VLDs, such as a lack of data governance and inconsistent data consent. Last but not least, we make recommendations on the implications of our findings to call for a more responsible curation and release of web-scraped VLDs that respect owners' data consent.

\clearpage
\bibliographystyle{plainnat}
\bibliography{references}

\section*{Acknowledgments} 
We would like to thank Christina Yeung for the thoughtful feedback on licensing, policy, and the writing revisions. This research is supported by the NSF Graduate Research Fellowship Program.

\clearpage
\beginsupplement
\section{Watermark Detection Details}
\label{sec:watermark-detection-details}
\subsection{Models} The off-the-shelf YoloV8 is finetuned on \citet{w6janfdataset} comprising 4,935 watermarked images by \citet{watermark-yolo-v8}. We finetune a pre-trained MobileViTv2~\citep{mehta2022separable} on the training split of \citet{pollano_watermarked_2019} comprising 12,510 and 12,477 images for watermarked and non-watermarked images. The pre-trained MobileViTv2~\citep{mehta2022separable} is loaded via Huggingface checkpoint \texttt{apple/mobilevitv2-1.0-imagenet1k-256}. We use Huggingface checkpoints for both Rolm-OCR and Gemma-3-12b-it, and we prompt the VLMs with: \textit{A watermark on an image is a deliberately embedded visual marker — often semi-transparent text, logos, or patterns — designed to assert ownership, deter unauthorized use, or signal authenticity. It can also be a form of a link, brand name, or author name at the top/bottom corner of the image. Does this image contain any watermark? If so, return the text of the watermark. Otherwise, return no in lowercase.}
\subsection{Compute Resources} All model training and evaluation use 2 Nvidia A100 GPUs.

\section{Copyright Notice Search Pattern}
\label{sec:copyright-notice-search-pattern}
\begin{figure}[!htbp]
    \centering
    \includegraphics[width=0.8\textwidth]{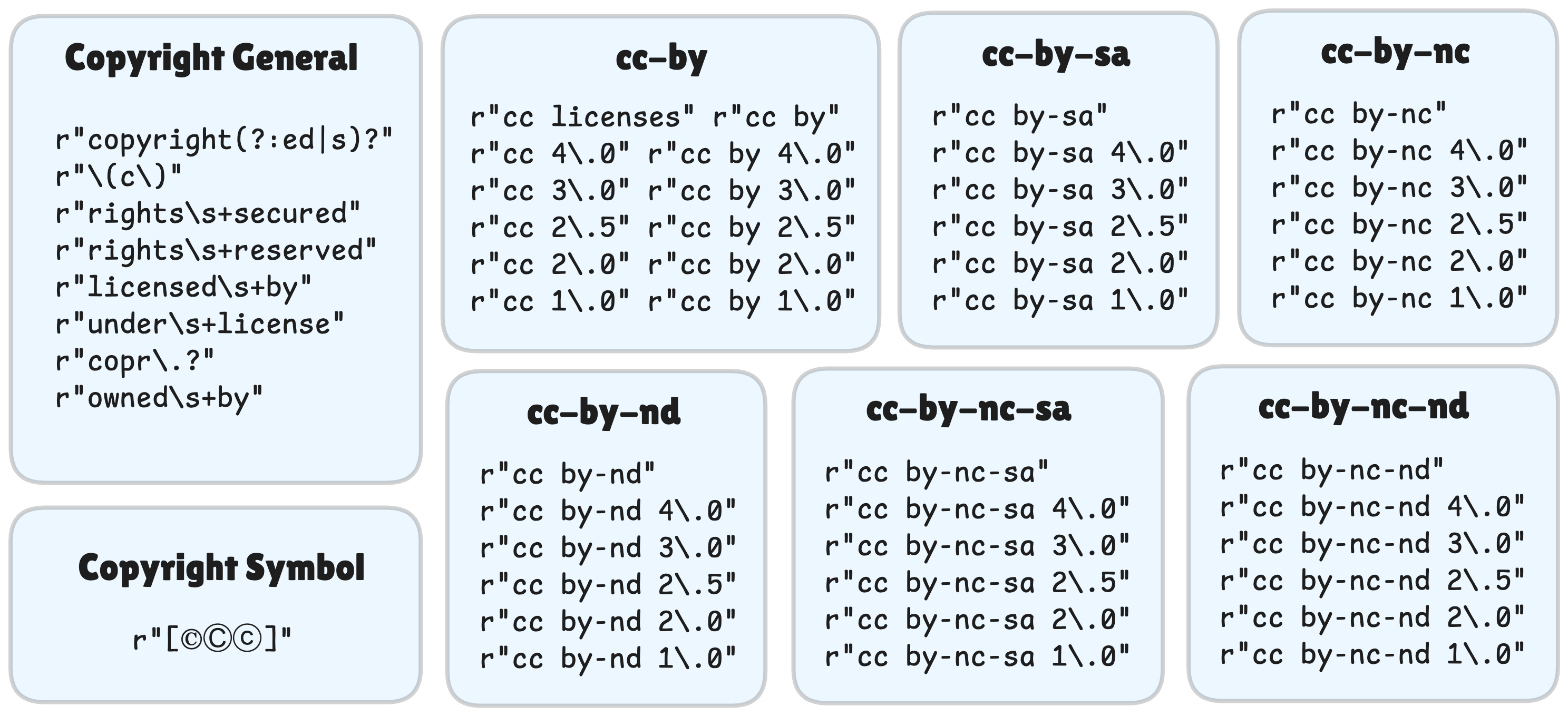}
    \caption{Regular expression search patterns used to source copyright notice in samples' captions and OCR-extracted texts.}
    \label{fig:copyright_search_pattern}
\end{figure}
\begin{figure}[!htbp]
    \centering
    \begin{subfigure}[t]{0.48\textwidth}
        \centering
        \includegraphics[width=\textwidth]{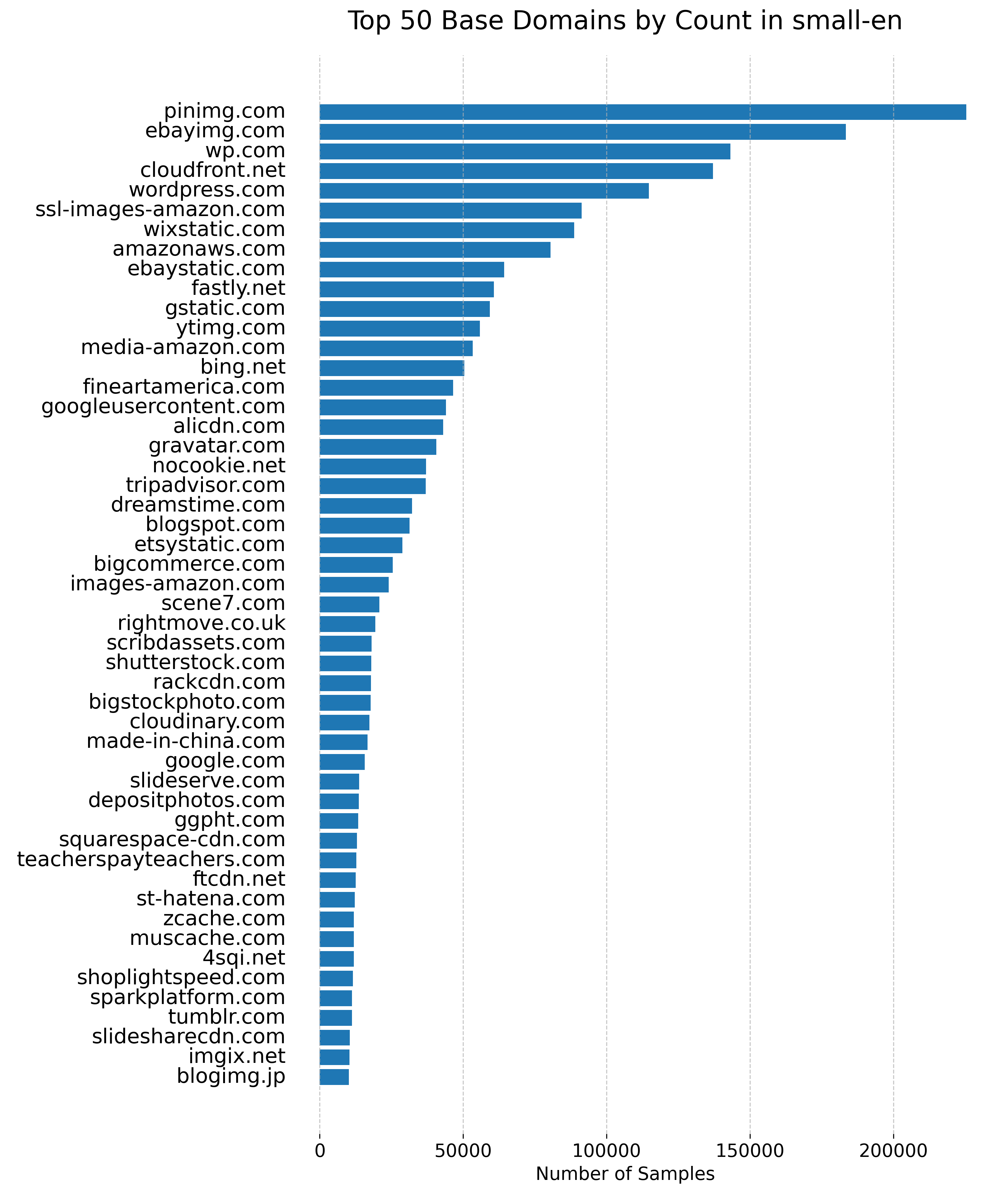}
        \caption{Top 50 base domains in \texttt{small-en} by sample counts.}
        \label{fig:top50-small-en}
    \end{subfigure}
    \hfill
    \begin{subfigure}[t]{0.48\textwidth}
        \centering
        \includegraphics[width=\textwidth]{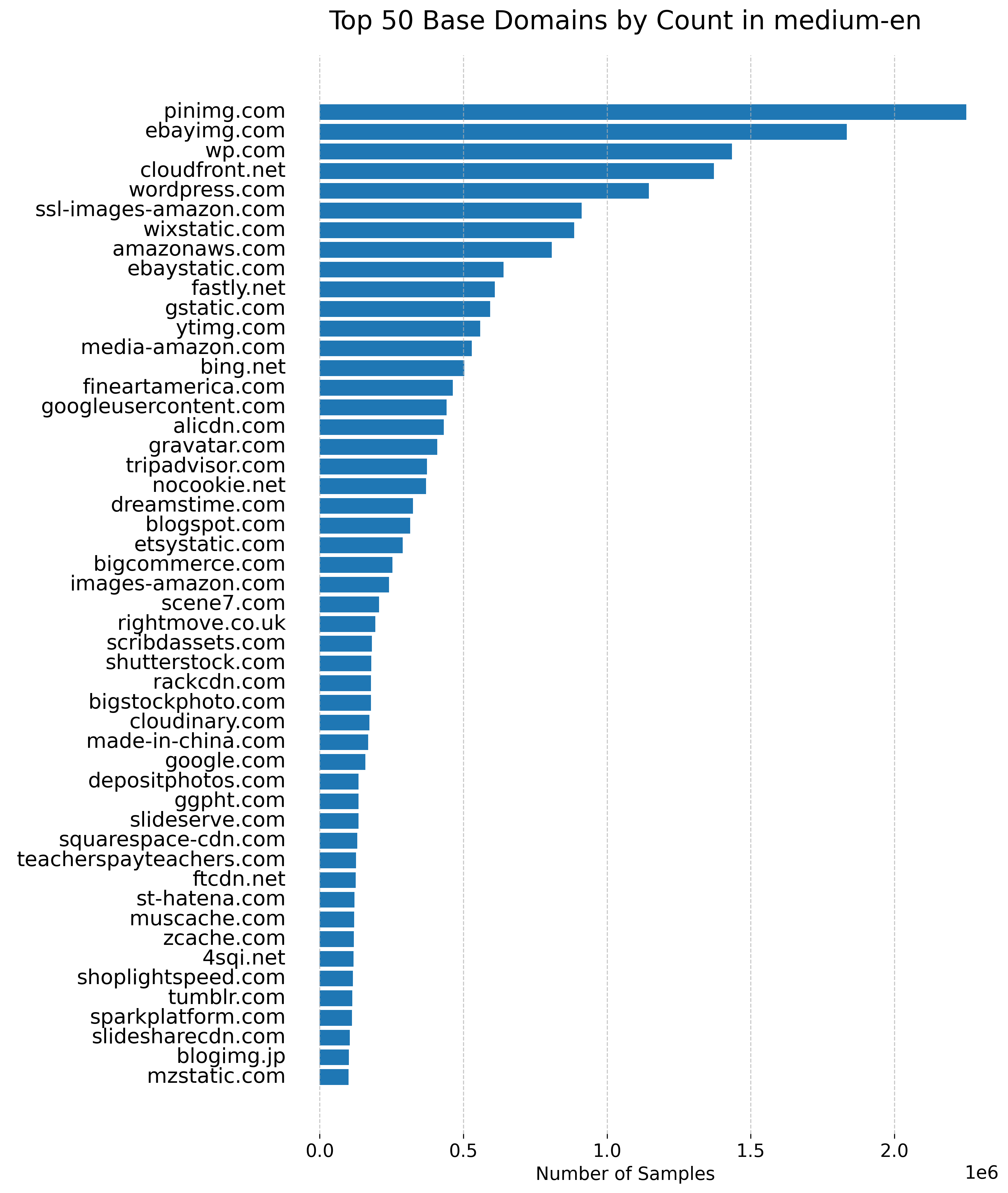}
        \caption{Top 50 base domains in \texttt{medium-en} by sample counts.}
        \label{fig:top50-medium-en}
    \end{subfigure}
    \caption{Distribution of the top 50 base domains in the \texttt{small-en} and \texttt{medium-en} splits of CommonPool. We observe the top 50 base domains only differ by one, where \texttt{small-en} has \texttt{imgix.net} and \texttt{medium-en} has \texttt{mzstatic.com}.}
    \label{fig:top50-base-domains}
\end{figure}
The full copyright notice search patterns are illustrated in Figure~\ref{fig:copyright_search_pattern}. Each category has multiple regular expression patterns. We find samples that have at least one match for any regular expression in the list. For ``Copyright General,'' we include commonly used patterns to claim copyright. For ``Copyright Symbol,'' we include three encoding variants of copyright symbols for better capture.  For the Creative Commons, we search for all 6 license types under Creative Commons, including the past versions.

\begin{table}
    \centering
    \begin{tabular}{ll}
        \toprule
        \textbf{Attribute} & \textbf{Types} \\
        \midrule
        Category & 
        \chipbox[colback=blue!10]{Marketplace} 
        \chipbox[colback=green!15]{CDN provider} \\[2pt]
        & \chipbox[colback=orange!20]{Blog} 
        \chipbox[colback=violet!15]{Website hosting} \\[2pt]
        & \chipbox[colback=cyan!10]{Stock photo} \\[2pt]
        & \chipbox[colback=gray!15]{Content-sharing community} \\[2pt] \midrule

        License Type & 
        \chipbox[colback=blue!10]{Personal/Noncommercial/Research} \\[2pt]
        & \chipbox[colback=green!15]{Conditional commercial use} \\[2pt]
        & \chipbox[colback=orange!20]{Open/Unrestricted commercial use} \\[2pt]
        & \chipbox[colback=violet!15]{Not applicable} \\[2pt] \midrule

        Scraping Policy & 
        \chipbox[colback=blue!10]{No scraping and AI conditionally} \\[2pt]
        & \chipbox[colback=green!15]{No scraping and AI} \chipbox[colback=orange!20]{No scraping} \\[2pt]
        & \chipbox[colback=violet!15]{Not mentioned} \\
        \bottomrule
    \end{tabular}
    \caption{Annotation schema for each attribute considered for web-domain-level characteristics. CDN Provider refers to third-party providers of content delivery network (CDN) as a service, such as Amazon Web Services.}
    \label{tab:annotation-schema}
\end{table}

\section{Terms of Service Analysis Codebook}
\label{sec:tos-codebook}
There are three attributes we annotate for each web domain: (1) Category, (2) License, and (3) Scraping Policy. Table \ref{tab:annotation-schema} summarizes the types included in each attribute. The codebook finalized for each attribute and type is as follows:
\begin{enumerate}
    \item \textit{Category}
    \begin{itemize}
        \item \textbf{Marketplace (E-commerce)} -- Platforms where \textit{general} goods or services are bought and sold.
        \item \textbf{CDN Provider} -- Content Delivery Network \textit{providers} and \textit{services} that deliver web content to users based on geographic location. For instance, \texttt{alicdn} and \texttt{cloudfront.net} fall under this type. This type does not include CDN incorporated by specific and mappable entities for faster content delivery. For instance, Adobe has its own CDN web domain to deliver its content instead of serving others' content.
        \item \textbf{Website Hosting Service} -- Services providing infrastructure for websites to be hosted and accessible on the internet. For instance, \texttt{wixstatic.com} and \texttt{wp.com} fall under this type.
        \item \textbf{Blog Service} -- Platforms for users to publish blogs. For instance, \texttt{blogspot.com} falls under this category.
        \item \textbf{Stock Photo Platform} -- Platforms where \textit{image assets} are bought and sold, typically under licensing agreements. This type differs from \textbf{Marketplace (E-commerce)} in that the \textit{goods} are \textit{image assets} themselves.
        \item \textbf{Content-sharing Community Platform} -- Platforms for user-generated and community-purposed content, as opposed to transaction-based exchanges.
        \item \textbf{Other} -- Uncommon websites or services that don't fall under any previous category. For instance, \texttt{4sqi.com} offers location-intelligence information through its API.
    \end{itemize}

    \item \textit{License Type}
    \begin{itemize}
        \item \textbf{Personal/Noncommercial/Research Only} -- Use of content is limited to personal, research, or noncommercial contexts. Commercial use is explicitly prohibited.
        \item \textbf{Conditional Commercial Access} -- Commercial use is permitted under certain conditions, such as requiring permission, excluding third-party redistribution, or purchasing a membership/plan.
        \item \textbf{Open or Unrestricted Commercial Use} -- Commercial use is allowed without restriction; the content is considered public or under an open license.
        \item \textbf{Not Applicable} -- The website does not specify any licensing or restrictions, or the service itself has no ruling over the content it hosts.
    \end{itemize}

    \item \textit{Scraping Policy}
    \begin{itemize}
        \item \textbf{No scraping and AI} -- Explicitly prohibits scraping and AI for any content.
        \item \textbf{No scraping} -- Explicitly prohibits scraping, but no mention of AI.
        \item \textbf{No AI} -- Explicitly prohibits AI, but no mention of scraping.
        \item \textbf{No scraping and AI conditionally} -- Prohibits a part of the content from scraping and AI, or prohibits scraping and AI under certain conditions, such as the permission of robots.txt.
        \item \textbf{Not Mentioned} -- No explicit restrictions mentioned around scraping or AI in the Terms of Service.
    \end{itemize}
\end{enumerate}

\section{Robots.txt Full Results}
\label{sec:robots-txt-full-results}
In the top 50 web domains from \texttt{small-en} and \texttt{medium-en}, we observe 3218 and 3879 agents, respectively. These observations cover 1,126,876 and 11,556,755 samples in \texttt{small-en} and \texttt{medium-en}, respectively. In table~\ref{tab:full-robots-txt-small-en} and table~\ref{tab:full-robots-txt-medium-en}, we see a very similar robots.txt analysis where the medium scale has about 10 times the observations as the total set scales up by 10 times. The dark gray background indicates that the ``All Disallowed'' rate, relative to the number of observations, is greater than or equal to 80\%. We observe that the all AI-purposed robots have over 80\% \textit{All Disallowed} rates.

\begin{table}[!htbp]
\resizebox{\textwidth}{!}{
\begin{tabular}{lc cc cc cc}
\toprule
\multirow{2}{*}{\textbf{Agent}} & \multirow{2}{*}{Observed} & \multicolumn{2}{c}{\textit{All Disallowed}} & \multicolumn{2}{c}{\textit{Some Disallowed}} & \multicolumn{2}{c}{\textit{None Disallowed}} \\
 &  & Count & \% of observed & Count & \% of observed & Count & \% of observed \\ \midrule
  *All Agents* & 1,126,876 & 6,442 & 0.6\% & 1,014,576 & 90.0\% & 105,858 & 9.4\% \\ \midrule
  \rowcolor{gray!20} GPTBot \faRobot & 578,498 & 538,431 & 93.1\% & 40,028 & 6.9\% & 39 & 0.0\% \\
  $\ast$ & 475,139 & 18,595 & 3.9\% & 391,799 & 82.5\% & 64,745 & 13.6\% \\
  \rowcolor{gray!20} CCBot \faRobot & 353,324 & 313,920 & 88.8\% & 39,365 & 11.1\% & 39 & 0.0\% \\
  \rowcolor{gray!20} Bytespider \faRobot & 301,344 & 262,029 & 87.0\% & 39,274 & 13.0\% & 41 & 0.0\% \\
  googlebot-image & 224,268 & 0 & 0.0\% & 224,166 & 100.0\% & 102 & 0.0\% \\
  \rowcolor{gray!20} claudebot \faRobot & 224,200 & 224,199 & 100.0\% & 1 & 0.0\% & 0 & 0.0\% \\
  \rowcolor{gray!20} Google-Extended \faRobot & 219,512 & 180,111 & 82.1\% & 39,367 & 17.9\% & 34 & 0.0\% \\
  \rowcolor{gray!20} SentiBot & 219,365 & 180,086 & 82.1\% & 39,274 & 17.9\% & 5 & 0.0\% \\
  Baiduspider & 204,497 & 35,762 & 17.5\% & 168,716 & 82.5\% & 19 & 0.0\% \\
  FacebookBot & 183,430 & 144,102 & 78.6\% & 39,288 & 21.4\% & 40 & 0.0\% \\
  omgili & 183,405 & 144,107 & 78.6\% & 39,274 & 21.4\% & 24 & 0.0\% \\
  Amazonbot & 183,399 & 144,070 & 78.6\% & 39,297 & 21.4\% & 32 & 0.0\% \\
  omgilibot & 183,118 & 143,820 & 78.5\% & 39,274 & 21.4\% & 24 & 0.0\% \\
  Googlebot-Image & 180,355 & 32 & 0.0\% & 168,841 & 93.6\% & 11,482 & 6.4\% \\
  \rowcolor{gray!20} Bingbot & 142,854 & 142,668 & 99.9\% & 40 & 0.0\% & 146 & 0.1\% \\
  Mediapartners-Google* & 59,654 & 0 & 0.0\% & 0 & 0.0\% & 59,654 & 100.0\% \\
  GoogleContextual & 59,231 & 0 & 0.0\% & 59,231 & 100.0\% & 0 & 0.0\% \\
  Twitterbot & 52,649 & 6 & 0.0\% & 40,463 & 76.9\% & 12,180 & 23.1\% \\
  bingbot & 49,452 & 7 & 0.0\% & 49,270 & 99.6\% & 175 & 0.4\% \\
  \rowcolor{gray!20} ClaudeBot \faRobot & 38,108 & 37,979 & 99.7\% & 91 & 0.2\% & 38 & 0.1\% \\
  \rowcolor{gray!20} Applebot-Extended \faRobot & 37,797 & 37,710 & 99.8\% & 55 & 0.1\% & 32 & 0.1\% \\
  \rowcolor{gray!20} PetalBot & 36,696 & 36,647 & 99.9\% & 1 & 0.0\% & 48 & 0.1\% \\
  \rowcolor{gray!20} magpie-crawler & 36,333 & 36,332 & 100.0\% & 0 & 0.0\% & 1 & 0.0\% \\
  applebot & 36,269 & 0 & 0.0\% & 36,222 & 99.9\% & 47 & 0.1\% \\
  AdsBot-Google & 28,599 & 25 & 0.1\% & 28,469 & 99.5\% & 105 & 0.4\% \\
  Yandex & 15,974 & 401 & 2.5\% & 15,552 & 97.4\% & 21 & 0.1\% \\
  facebookexternalhit & 15,678 & 7 & 0.0\% & 37 & 0.2\% & 15,634 & 99.7\% \\
  AdIdxBot & 12,927 & 0 & 0.0\% & 12,905 & 99.8\% & 22 & 0.2\% \\
  Googlebot & 12,303 & 26 & 0.2\% & 392 & 3.2\% & 11,885 & 96.6\% \\
  Pinterestbot & 11,950 & 7 & 0.1\% & 11,891 & 99.5\% & 52 & 0.4\% \\
  ia\_archiver & 4,983 & 131 & 2.6\% & 4,695 & 94.2\% & 157 & 3.2\% \\
  \rowcolor{gray!20} anthropic-ai \faRobot & 1,739 & 1,689 & 97.1\% & 14 & 0.8\% & 36 & 2.1\% \\
  \rowcolor{gray!20} ImagesiftBot & 1,636 & 1,592 & 97.3\% & 1 & 0.1\% & 43 & 2.6\% \\
  \rowcolor{gray!20} meta-externalagent \faRobot & 1,414 & 1,398 & 98.9\% & 2 & 0.1\% & 14 & 1.0\% \\
  \rowcolor{gray!20} PerplexityBot & 1,409 & 1,223 & 86.8\% & 138 & 9.8\% & 48 & 3.4\% \\
  \rowcolor{gray!20} MJ12bot & 1,033 & 982 & 95.1\% & 5 & 0.5\% & 46 & 4.5\% \\
\bottomrule
\end{tabular}
}
\caption{Top results from robots.txt analysis for \texttt{small-en} scale's top 50 \textit{base domains}, accounting for 96,436 attempted \textit{full domains}, 81,273 successful robots.txt, and 1,126,876 samples. The full list of agents is not shown for conciseness. In this table, we only show agents with over 1,000 sample observations. The dark gray background highlights agents that have over 80\% ``All Disallowed'' rate. For each agent, the number of observed cases is broken down by the number and percentage (relative to observed) of cases where all, some, or none were disallowed. ``All Agents'' row refers to an aggregation of all agents found in all the examined robots.txt. The aggregation rule is as follows: If for all agents, a robots.txt has \textit{All Disallowed}, then the decision is \textit{All Disallowed}. If for any agent in all agents, a robots.txt has \textit{All Disallowed} or \textit{Some Disallowed}, then a robots.txt has \textit{Some Disallowed}. Otherwise, it has \textit{None Disallowed}.}
\label{tab:full-robots-txt-small-en}
\end{table}

\begin{table}[!htbp]
\resizebox{\textwidth}{!}{
\begin{tabular}{lc cc cc cc}
\toprule
\multirow{2}{*}{\textbf{Agent}} & \multirow{2}{*}{Observed} & \multicolumn{2}{c}{\textit{All Disallowed}} & \multicolumn{2}{c}{\textit{Some Disallowed}} & \multicolumn{2}{c}{\textit{None Disallowed}} \\
 &  & Count & \% of observed & Count & \% of observed & Count & \% of observed \\
\midrule
  *All Agents* & 11,556,755 & 65,886 & 0.6\% & 10,521,922 & 91.0\% & 968,947 & 8.4\% \\ \midrule
  \rowcolor{gray!20} GPTBot \faRobot & 5,781,111 & 5,378,225 & 93.0\% & 402,335 & 7.0\% & 551 & 0.0\% \\
  $\ast$ & 5,039,780 & 186,668 & 3.7\% & 4,296,202 & 85.2\% & 556,910 & 11.1\% \\
  \rowcolor{gray!20} CCBot \faRobot & 3,532,300 & 3,136,474 & 88.8\% & 395,388 & 11.2\% & 438 & 0.0\% \\
  \rowcolor{gray!20} Bytespider \faRobot & 3,014,323 & 2,619,564 & 86.9\% & 394,413 & 13.1\% & 346 & 0.0\% \\
  googlebot-image & 2,239,424 & 0 & 0.0\% & 2,238,505 & 100.0\% & 919 & 0.0\% \\
  \rowcolor{gray!20} claudebot \faRobot & 2,238,757 & 2,238,756 & 100.0\% & 1 & 0.0\% & 0 & 0.0\% \\
  \rowcolor{gray!20} Google-Extended \faRobot & 2,203,460 & 1,807,713 & 82.0\% & 395,391 & 17.9\% & 356 & 0.0\% \\
  \rowcolor{gray!20} SentiBot & 2,201,585 & 1,807,137 & 82.1\% & 394,404 & 17.9\% & 44 & 0.0\% \\
  Baiduspider & 2,040,055 & 357,497 & 17.5\% & 1,682,293 & 82.5\% & 265 & 0.0\% \\
  Amazonbot & 1,838,597 & 1,443,521 & 78.5\% & 394,671 & 21.5\% & 405 & 0.0\% \\
  FacebookBot & 1,837,341 & 1,442,411 & 78.5\% & 394,581 & 21.5\% & 349 & 0.0\% \\
  omgili & 1,836,966 & 1,442,343 & 78.5\% & 394,410 & 21.5\% & 213 & 0.0\% \\
  omgilibot & 1,835,306 & 1,440,691 & 78.5\% & 394,406 & 21.5\% & 209 & 0.0\% \\
  Googlebot-Image & 1,798,281 & 75 & 0.0\% & 1,683,359 & 93.6\% & 114,847 & 6.4\% \\
  \rowcolor{gray!20} Bingbot & 1,431,194 & 1,429,300 & 99.9\% & 356 & 0.0\% & 1,538 & 0.1\% \\
  Mediapartners-Google* & 597,643 & 1 & 0.0\% & 0 & 0.0\% & 597,642 & 100.0\% \\
  GoogleContextual & 592,649 & 0 & 0.0\% & 592,649 & 100.0\% & 0 & 0.0\% \\
  Twitterbot & 529,106 & 41 & 0.0\% & 407,550 & 77.0\% & 121,515 & 23.0\% \\
  bingbot & 498,101 & 66 & 0.0\% & 496,491 & 99.7\% & 1,544 & 0.3\% \\
  ia\_archiver & 434,741 & 1,339 & 0.3\% & 431,807 & 99.3\% & 1,595 & 0.4\% \\
  \rowcolor{gray!20} ClaudeBot \faRobot & 384,024 & 382,664 & 99.6\% & 949 & 0.2\% & 411 & 0.1\% \\
  \rowcolor{gray!20} Applebot-Extended \faRobot & 380,818 & 380,218 & 99.8\% & 335 & 0.1\% & 265 & 0.1\% \\
  \rowcolor{gray!20} PetalBot & 370,568 & 370,078 & 99.9\% & 18 & 0.0\% & 472 & 0.1\% \\
  \rowcolor{gray!20} magpie-crawler & 366,942 & 366,927 & 100.0\% & 4 & 0.0\% & 11 & 0.0\% \\
  applebot & 366,462 & 0 & 0.0\% & 365,972 & 99.9\% & 490 & 0.1\% \\
  AdsBot-Google & 287,314 & 365 & 0.1\% & 285,986 & 99.5\% & 963 & 0.3\% \\
  Yandex & 160,006 & 2,901 & 1.8\% & 156,800 & 98.0\% & 305 & 0.2\% \\
  facebookexternalhit & 158,068 & 137 & 0.1\% & 310 & 0.2\% & 157,621 & 99.7\% \\
  AdIdxBot & 129,314 & 1 & 0.0\% & 129,124 & 99.9\% & 189 & 0.1\% \\
  Googlebot & 122,753 & 354 & 0.3\% & 3,677 & 3.0\% & 118,722 & 96.7\% \\
  Pinterestbot & 119,439 & 112 & 0.1\% & 118,796 & 99.5\% & 531 & 0.4\% \\
  \rowcolor{gray!20} anthropic-ai \faRobot & 16,224 & 15,728 & 96.9\% & 196 & 1.2\% & 300 & 1.8\% \\
  \rowcolor{gray!20} ImagesiftBot & 15,332 & 14,904 & 97.2\% & 18 & 0.1\% & 410 & 2.7\% \\
  \rowcolor{gray!20} meta-externalagent \faRobot & 14,945 & 14,790 & 99.0\% & 8 & 0.1\% & 147 & 1.0\% \\
  \rowcolor{gray!20} PerplexityBot & 14,413 & 12,513 & 86.8\% & 1,435 & 10.0\% & 465 & 3.2\% \\
  \rowcolor{gray!20} MJ12bot & 10,425 & 9,845 & 94.4\% & 41 & 0.4\% & 539 & 5.2\% \\
\bottomrule
\end{tabular}
}
\caption{Top results from robots.txt analysis for \texttt{medium-en} scale's top 50 \textit{base domains}, accounting for 434,498 attempted \textit{full domains}, 392,286 successful robots.txt, and 11,556,755 samples. The full list of agents is not shown for conciseness. In this table, we only show agents with over 10,000 sample observations. The dark gray background highlights agents that have over 80\% ``All Disallowed'' rate. For each agent, the number of observed cases is broken down by the number and percentage (relative to observed) of cases where all, some, or none were disallowed. ``All Agents'' row refers to an aggregation of all agents found in all the examined robots.txt. The aggregation rule is as follows: If for all agents, a robots.txt has \textit{All Disallowed}, then the decision is \textit{All Disallowed}. If for any agent in all agents, a robots.txt has \textit{All Disallowed} or \textit{Some Disallowed}, then a robots.txt has \textit{Some Disallowed}. Otherwise, it has \textit{None Disallowed}.}
\label{tab:full-robots-txt-medium-en}
\end{table}

\end{document}